\documentclass[preprint,prb,aps,showpacs,groupedaddress]{revtex4-1}
\usepackage{graphicx}
\usepackage{amssymb}
\usepackage{textcomp}
\usepackage{amsmath}
\usepackage[normalem]{ulem} 
\usepackage{soul} 
\def\be{\begin{equation}}
\def\ee{\end{equation}}
\def\bea{\begin{eqnarray}}
\def\eea{\end{eqnarray}}
\def\bdm{\begin{displaymath}}
\def\edm{\end{displaymath}}
\def\ba{\begin{array}}
\def\ea{\end{array}}

\begin{document}

\title{Friedel oscillation near a van Hove singularity in two-dimensional Dirac materials}

\author{Chi-Ken Lu}
\email{Lu49@ntnu.edu.tw}

\address{Physics Department, National Taiwan Normal University, Taipei 11677, Taiwan}

\date{\today}

\begin{abstract}

We consider Friedel oscillation in the two-dimensional Dirac materials when Fermi level is near the van Hove singularity. Twisted graphene bilayer and the surface state of topological crystalline insulator are the representative materials which show low-energy saddle points that are feasible to probe by gating. We approximate the Fermi surface near saddle point with a hyperbola and calculate the static Lindhard response function. Employing a theorem of Lighthill, the induced charge density $\delta n$ due to an impurity is obtained and the algebraic decay of $\delta n$ is determined by the singularity of the static response function. Although a hyperbolic Fermi surface is rather different from a circular one, the static Lindhard response function in the present case shows a singularity similar with the response function associated with circular Fermi surface, which leads to the $\delta n\propto R^{-2}$ at large distance $R$. The dependences of charge density on the Fermi energy are different. Consequently, it is possible to observe in twisted graphene bilayer the evolution that $\delta n\propto R^{-3}$ near Dirac point changes to $\delta n\propto R^{-2}$ above the saddle point. Measurements using scanning tunnelling microscopy around the impurity sites could verify the prediction.




\end{abstract}


\maketitle

\section{introduction}

Saddle points quite often appear in band structure of two-dimensional crystals such as cuprate, and a logarithmic divergence, the van Hove singularity,~\cite{vanHove,vHS1,vHS2,vHS3} in density of state (DOS) is derived from the hyperbolic band $E({\bf k})\propto k_x^2-k_y^2$ near it. In addition, Fermi surfaces of different pockets in the Brillouin zone touch each other right at the saddle point. Based on weak-coupling theory, the divergent DOS was argued to dramatically raise the transition temperature of superconductivity as chemical potential moves toward the saddle point.~\cite{Tsuei,highTC} Recently, the possibility of superconducting instability in doped graphene~\cite{dopedG} was investigated using Kohn-Luttinger theory~\cite{Gonzalez0} and renormalization group analysis.~\cite{Nandkishore} From practical point of view, however, graphene with Fermi level at several electron volts away from charge neutral point requires a significant amount of doping to achieve the desired state, which is difficult with current gating technology.~\cite{dopedG} Thus, the twisted graphene bilayer (tGB)~\cite{tBG0} and the topological crystalline insulator (TCI)~\cite{TCI1} have drawn much attention due to the relatively low-energy van Hove singularity located at less than a hundred mili electron volts (meV) away from the Dirac point. More recently, phosphorene, a single layer of black phosphorus, with a saddle point near the Fermi energy is also an interesting candidate material.~\cite{P0,P1,HLin}

With the rapidly increasing research activities focusing on electronic~\cite{tBGa,Shallcross,Brey} and magnetic~\cite{tBG1,tBGKorea,Bis,tBGMoon,tBGLu1,tBGLu2,FuTCI_PRB} properties of saddle points in two-dimensional Dirac materials, it is important to explore and predict measurable quantities which have root in the hyperbolic band near the saddle point. One particular aspect is to observe the carrier density oscillation due to a localised impurity or defect, which can be implemented with the technique of scanning tunnelling microscopy (STM).~\cite{STM} The oscillation, first predicted by Friedel~\cite{Friedel} and thus named Friedel oscillation (FO), is a unique consequence of the sharp Fermi surfaces of metals. In the process of elastic scattering by an impurity, the largest momentum transfer acquired by an electron is $2k_F$, which corresponds to the nesting vector of isotropic Fermi surface. The long-range FO is also determined by the electron's wave functions around the Fermi surface. For two-dimensional electron gas with a parabolic band, Stein showed that the local charge density $\delta n(R)\approx \sin(2k_FR)/R^2$ for $k_FR\gg1$ away from the single impurity.~\cite{Stern} In doped graphene and other two-dimensional Dirac materials, electrons near the Fermi surface enclosing a Dirac point are described by two-component wavefunctions. The linear dispersion near the Dirac point results in the Berry phase of $\pi$ and, consequently, a distinct dependence of $1/R^{3}$ in FO, which is accounted for by the lack of backward scattering near the Fermi surface.~\cite{Ando,DHLin,Falko,Guinea,Hwang,Herb} 

\begin{figure}
\input{epsf}
\includegraphics[width=0.44\textwidth]{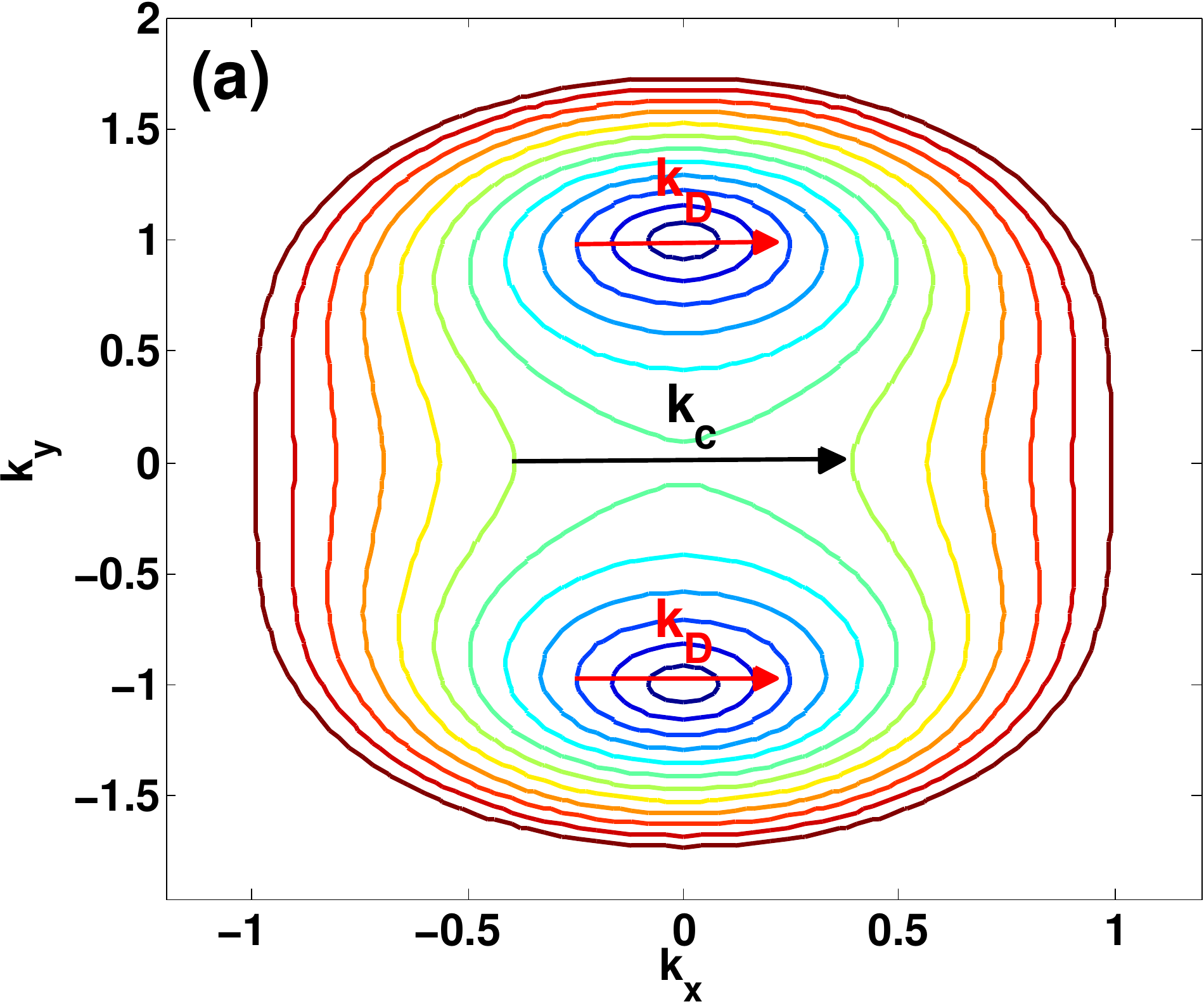}
\includegraphics[width=0.45\textwidth]{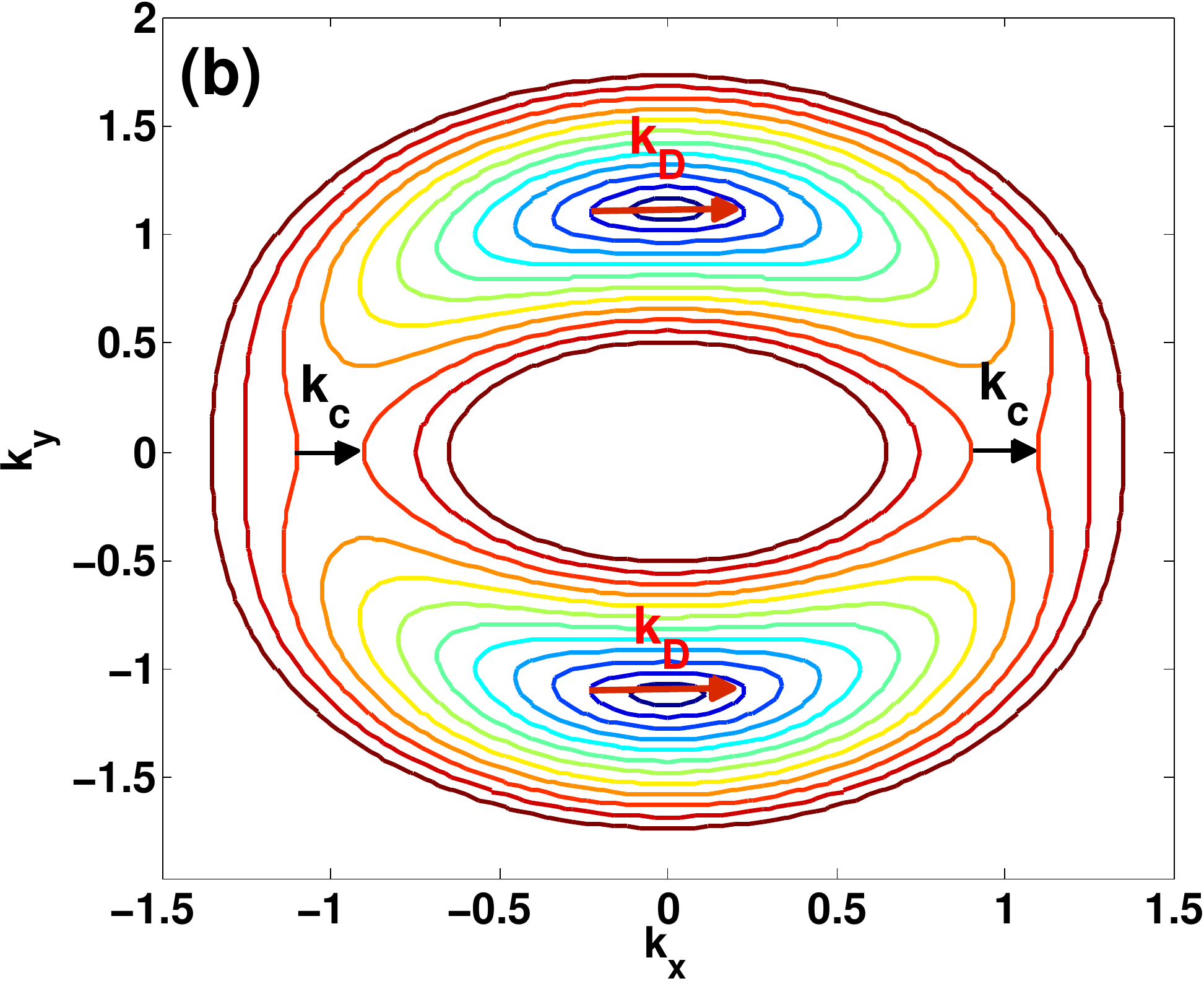}
\caption{ (color online) Constant energy contours of the twisted graphene bilayer in panel (a) and the surface state of topological crystalline insulator in panel (b). The red arrows $\bf k_D$ in both panels represent the nesting vectors associated with the Fermi surface enclosing a Dirac point. When the Fermi level is raised above the saddle point, the dark arrows $\bf k_c$ connect different patches of the Fermi surface in both panels. The direction of $k_y$ in panel (a) is parallel with the line joining the two adjacent Dirac points from opposite layers in tGB. In panel (b) $k_x$ ($k_y$) is parallel with $\bar\Gamma\bar X_1$ ($\bar\Gamma\bar X_2$) defined in Ref.~\onlinecite{FuTCI_PRB}.}\label{contour}
\end{figure}

In contrast to the low-energy Fermi surfaces enclosing Dirac point, to the best of our knowledge, there has been little or none investigation regarding the FO in the case when Fermi level is close to or at the saddle point. In this paper we treat tGB and the surface states of TCI as examples and study the FO when Fermi level is near the saddle points. Using the band structures obtained in simple models, the evolutions of Fermi surfaces in tGB and TCI can be seen in Fig.~\ref{contour}. The colored arrows labelled by $\bf k_D$ and $\bf k_c$ represent the nesting vector when the Fermi level is near the Dirac point and saddle point, respectively. Besides, we need to evaluate the Lindhard response function  
\be
	\Pi({\bf q},\Omega) = \int\frac{d{\bf k}}{4\pi^2}\frac{f_{\bf k+q}-f_{\bf k}}{\hbar\Omega-E_{\bf k+q}+E_{\bf k}+i\delta} F(\bf k, \bf k+q)\:,\label{lindhard}
\ee with form factor $F$ associated with the overlap between the two states connected by the momentum transfer $\bf q$. In the static limit $\Omega\rightarrow 0$, the Lindhard function has a singularity near the nesting vector, which is related to Kohn anomaly.~\cite{Kohn} Technically, FO, represented by the induced charge density $\delta n(R)$, is encoded in the Fourier transformation of the dielectric function ${\epsilon(\bf q)}$ with respect to the momentum transfer $\bf q$. While $\delta n(R)$ is vanishingly small for large $R$, the singularity appeared in $\Pi$ at the nesting vector, $\bf q=k_D$ or $\bf q=k_c$, gives rise to the desired FO. For this purpose and given the Lighthill theorem~\cite{lighthill} which deals with the asymptotic behaviour of Fourier transform, it suffices to obtain the leading term in $\delta n(R)$ from the analytic expression of $\Pi(\bf q)$ in the vicinity of nesting vector. The circular and hyperbolic Fermi surfaces associated with two-dimensional electron gas (2DEG) and saddle point are different in the geometrical sense, and we find rather different functional forms of Lindhard response functions in the two cases. However, the leading singularity behaviours in both cases are found to be of the same order, which leads to similar oscillatory $\delta n \propto R^{-2}$. For the special case at saddle point, we argue that the oscillatory term in $\delta n$ disappears despite that the bare $\Pi$ does contain a singularity.

The paper has the following organizations. We introduce the band structure of tGB and surface states in TCI with two simplified models in Sec.~\ref{model} and discuss the wavefunctions associated with the states along the Fermi surface. We also argue that the form factor $F$ factor can be dropped in the calculation of $\Pi$ in Sec.~\ref{pi} for the hyperbolic Fermi surface near the saddle points. In Sec.~\ref{fo}, the analytical properties of $\Pi$ are then used in obtaining the $\delta n$, and a final discussion is given in Sec.~\ref{dis}. 

\section{Saddle point in Model band structure}\label{model}

We first consider tGB with a two-band model reproducing the band structure of a pair of Dirac points and a pair of saddle points at opposite energies.~\cite{tBG1,nematic2} In real material, there are other saddle points in the Brillouin zone but we neglect them for the moment. In terms of the Pauli matrices $\sigma_x$ and $\sigma_y$ representing the mixed layer-sublattice characters of tGB, the Hamiltonian reads
\be
	H=\frac{1}{2m}[(k_x^2-k_y^2+K^2)\sigma_x+(2k_xk_y)\sigma_y]\:,
\ee from which the zero-energy Dirac points are observed to locate at the pair of momenta $(0,\pm K)$. The saddle points are found at the origin with energy of $\pm K^2/2m$. The term $(K^2/2m)\sigma_x$ breaks the rotational symmetry of the otherwise isotropic quadratic band touching point. The constant-energy contours of spectrum $E_{\pm}=\pm\frac{1}{2m}\sqrt{k^4+2K^2(k_x^2-k_y^2)+K^4}$ are shown in the left panel of Fig.~\ref{contour}. For momentum close to the Dirac points $(0,\pm K)$, the Hamiltonian approximately reads $H\approx K(\mp\delta k_y\sigma_x\pm \delta k_x\sigma_y)/m$, respectively, with $\delta k_x$ and $\delta k_y$ small in comparison with $K$. The winding number is the same for the two Dirac points. Eigenfunctions corresponding to energy $\frac{K}{m}\delta k$ are labelled by $\Psi^D_{\delta {\bf k}}=(i\ e^{-i\phi})^T/\sqrt{2}$ with $\phi$ being the polar angle associated with $\delta\bf k$. Now we focus on the saddle point at positive energy $K^2/2m$. Near the origin, the spectrum is approximately $E\approx(K^2+k_x^2-k_y^2)/2m$, and the corresponding eigenfunction associated with $E=\frac{K^2}{2m}+\mu$ is written as, 
\be
	\Psi^{v}_{\bf k,\mu} \propto 
	\left(\begin{array}{cccc}
	1\\
	1\end{array}\right)+
	\left(\begin{array}{cccc}
	\frac{k_x^2-k_y^2}{K^2}\\
	\frac{2m\mu}{K^2}\end{array}\right)
	\:,
\ee where the first term is the eigenvector of $\sigma_x$ with eigenvalue 1. In contrast to the wavefunctions $\Psi^D$ which results in the vanishing form factor $F=\langle\Psi^D_{-\delta \bf k}|\Psi^D_{\delta\bf k}\rangle$ between the two states connected by the nesting vector $\bf k_D$ shown in Fig.~\ref{contour}, the form factor $F=\langle\Psi^v_{\bf k,\mu}|\Psi^v_{\bf k',\mu}\rangle$ between the two states $\bf k$ and $\bf k'=k+k_c$ is always nonzero. The vanishing of $F$ in the case of slightly doped graphene is essential to the evaluation of Lindhard function since it has been demonstrated that different order of singularity is generated in $\Pi$.~\cite{Herb,Hwang1} On the other hand, the nonzero form factor $F$ can be ignored to simplify the calculations if we are mainly concerned with the singular behaviour of $\Pi$ for $\bf q=k_c$.

The surface of TCI also consists of a pair of Dirac points at zero energy, but, in contrast with tGB, there are two saddle points at positive energy as shown in the right panel in Fig.~\ref{contour}. Following Ref.~\onlinecite{FuTCI_PRB}, the simplified four-band Hamiltonian, which neglects the anisotropic Fermi velocities,
\be
	\mathcal H({\bf k}) = k_x\sigma_y+k_y\sigma_x+\Delta\tau_x+\delta\sigma_x\tau_y\:,
\ee
results in the spectrum $E^2=k^2+\Delta^2+\delta^2\pm2\sqrt{k^2\Delta^2+k_y^2\delta^2}$. The saddle point band structure is found in $E(\Delta+k_x,k_y)=\delta+\frac{k_x^2}{8\delta}-\frac{k_y^2}{2\Delta^2/\delta}$ on the x-axis. The negative effective mass $m_y=-\Delta^2/\delta$ can be understood from rewriting the four-band Hamiltonian $\mathcal H(\Delta+k_x,k_y)$ in the following form,
\be
	\mathcal H = \left(\begin{array}{cccc}
	-k_x & -i\delta & k_y & 0\\
	i\delta & k_x & 0 & k_y \\
	k_y & 0 & 2\Delta+k_x & -i\delta \\
	0 & k_y & i\delta & -2\Delta-k_x\end{array}\right)\:.
\ee The four levels with energy $\pm\sqrt{4\Delta^2+\delta^2}$ and $\pm\delta$ are obtained at the momentum $(\Delta,0)$. Considering the second-order perturbations from the levels of $E=\pm\sqrt{4\Delta^2+\delta^2}$ to the level of $E=\delta$, the above off-diagonal term proportional to $k_y$ gives rise to the desired negative mass. In addition, the wavefucntions $\Psi^v$ near the van Hove singularity at $E=\delta$ has the dominant component in $(1\ i\ 0\ 0)^T/\sqrt{2}$. Therefore, based on the same reasoning, it is valid to ignore the form factor $F$ when evaluating the Lindhard function when chemical potential is near the saddle point.


\section{Static Lindhard function}\label{pi}

\begin{figure}
\input{epsf}
\includegraphics[width=0.45\textwidth]{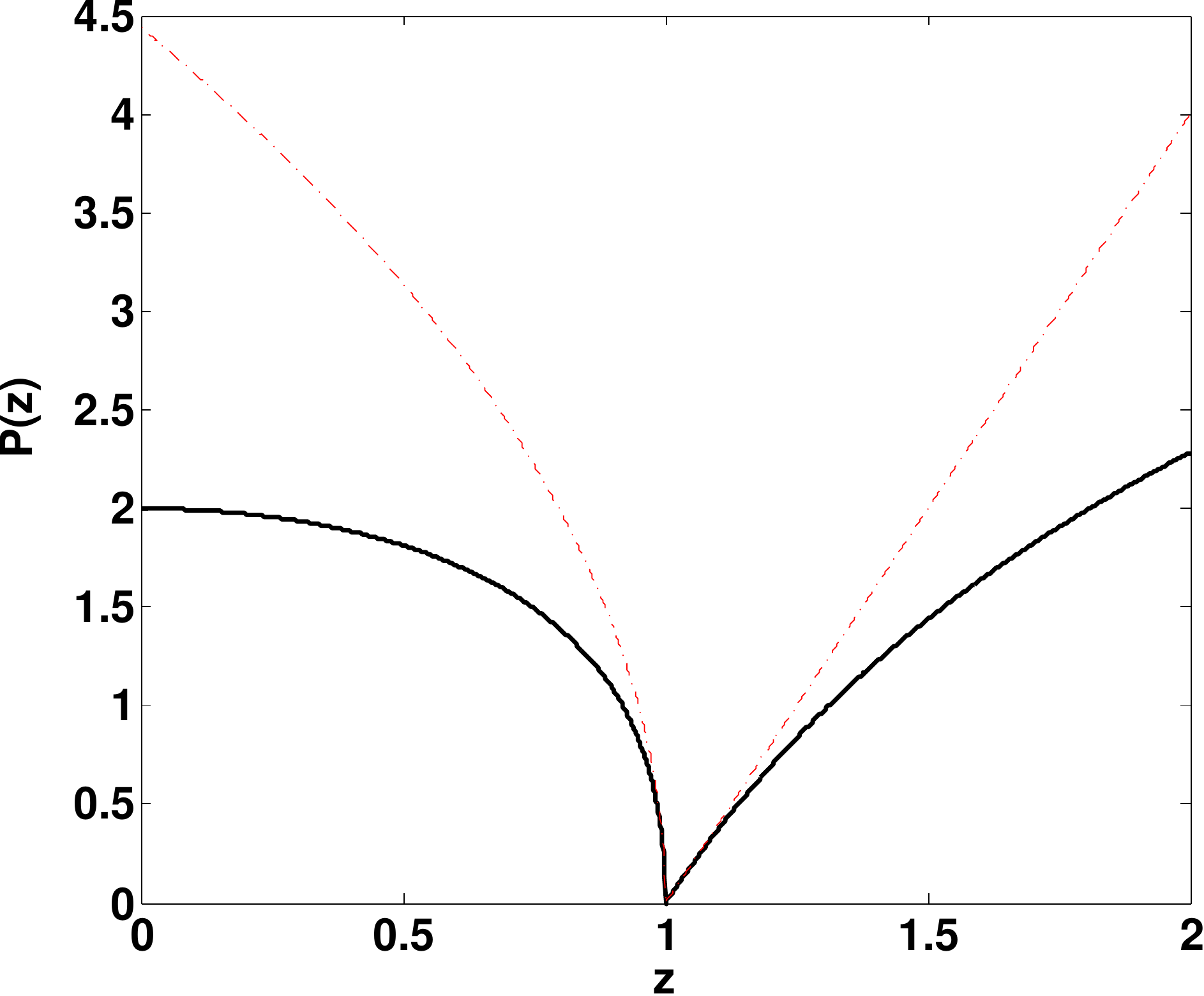}
\caption{(color online) $P(z)$ from Eq.~\ref{p1} and \ref{p2} is shown in dark lines. The red dashed lines represent the approximate forms near the cusp at $z=1$.}\label{Pi}
\end{figure}

In this section we evaluate $\Pi$ in the static case and focus on the Fermi surfaces close to or coinciding with the saddle point. We argue in previous section that the suppression of backscattering does not occur for momentum near the saddle point so the form factor $F$ in Eq.~\ref{lindhard} can be dropped. The Fermi surfaces near the saddle point can be specified by the solutions $\frac{k_x^2}{2m_x}-\frac{k_y^2}{2m_y}-\mu=0$, and the nesting vector $(k_c,0)$ is given by $k_c=2\sqrt{2m_x\mu}$. The critical Fermi surface is characterized by $k_c=0$. $\Pi$ depends on the vector $\bf q$ as well as the Fermi energy specified by $\mu$. Set $\Omega=0$ and rescale the momentum $\bf k$, 
\be
	\Pi^{\text{vHs}}({\bf q}) 
	= \frac{D}{\pi}\int_{-\Lambda}^{\Lambda}\int_{-\Lambda_x}^{\Lambda_x} \frac{dk_xdk_y}{2q_xk_x-2q_yk_y+q_x^2-q_y^2}+(\bf q\rightarrow-\bf q)\:,
\ee where a superscript $v$ is designated for its use near the van Hove singularity and $D\equiv m/2\pi\hbar^2$ stands for the density of state for a parabolic band with effective mass $m=\sqrt{m_xm_y}$. Besides, we have imposed a momentum cutoff $\Lambda$, and $\Lambda_x\equiv\sqrt{k_y^2+k_c^2/4}$. For ${\bf q}=q_x\hat x$, 
we proceed with integration by part and carefully deal with the sign of argument of logarithmic function, which leads to 
\be
\begin{split}
	\Pi^{\text{vHs}}(q_x\hat x)&=\frac{2D\Lambda}{\pi q_x}\ln\frac{\sqrt{\Lambda^2+k_c^2/4}+q_x/2}{\sqrt{\Lambda^2+k_c^2/4}-q_x/2}\\
	&+\frac{4D}{\pi q_x^2}\int_0^{\Lambda}dk_yk_y^2\left(
	\frac{-2}{\sqrt{k_y^2+k_c^2/4}}+\frac{1}{\sqrt{k_y^2+k_c^2/4}+q_x/2}+\frac{1}{\sqrt{k_y^2+k_c^2/4}-q_x/2}
	\right)\:.
\end{split}
\ee For the case of $k_c=0$, which corresponds to the critical Fermi surface coinciding with saddle point, it is easy to show, assuming $\Lambda\gg q_x$, that 

\be
	\Pi^{\text{vHs}}_0(q_x\hat x)=\frac{D}{\pi}(2+\ln\frac{4\Lambda^2}{q_x^2})\:,\label{sp_pi}
\ee which is identical with the corresponding expression in Ref.~\onlinecite{Gonzalez1}. We purposely attach a subscript $0$ to it, emphasising that $\Pi^{\text{vHs}}_0$ is associated with the Fermi surface of $\mu=0$. For positive $k_c$ without loss of generality, defining $z\equiv q_x/k_c$ and $\lambda\equiv2\Lambda/k_c$, we find
\be
	\Pi^{\text{vHs}}(z)=\frac{2D}{\pi}+\frac{D}{\pi}\int_0^{\Theta}d\theta  \left\{2\sec\theta+\frac{1-z^2}{z}\left[\frac{1}{z\cos\theta+1} + \frac{1}{z\cos\theta-1}\right] \right\}\:,
\ee  with $\Theta\approx\pi/2$ for large cutoff $\Lambda\gg k_c$. Now we arrive at the key result of the present paper,
\be
	\Pi^{\text{vHs}}(z)=\frac{D}{\pi}\left[2+\ln(4\lambda^2)-P(z)\right]\:,
\ee in which the second term diverges logarithmically as $k_c\rightarrow0$. The singularity of $\Pi$ is then encoded in $P$, which has an expression depending on whether $z$ is less or greater than unity. Using the Appendix, we obtain, for $z<1$, 
\be
	P(z) = \frac{2\sqrt{1-z^2}}{z}\left[
	\arctan\sqrt{\frac{1+z}{1-z}}-\arctan\sqrt{\frac{1-z}{1+z}}
	\right]\:,\label{p1}
\ee from which $P(z=0)=2$ is deduced. 
For $z>1$,
\be
	P(z) = \frac{2\sqrt{z^2-1}}{z}\ln\frac{\sqrt{z+1}+\sqrt{z-1}}{\sqrt{z+1}-\sqrt{z-1}}\:.\label{p2}
\ee The function $P$ is plotted numerically in Fig.~\ref{Pi}. A cusp appears at $z=1$, around which $\Pi^{\text{vHs}}$ has a discontinuous derivative. More precisely, $P$ has the following expansion near $z=1$, 
\[
    P(z)\approx 
\begin{cases}
    \pi\sqrt{2(1-z)},& \text{if } z< 1\\
    4(z-1),              & \text{if } z>1.
\end{cases}
\]

For comparison, we list the corresponding Lindhard function associated with 2DEG,~\cite{Stern} 
\be
	\Pi^{\text{2DEG}}(z\hat q) = D\left(
	1-\Theta(z-1)\sqrt{1-\frac{1}{z^2}}
	\right)\:,
\ee from which one can easily see the singular term proportional to $\Theta(z-1)\sqrt{z-1}$ is the leading contribution responsible for the FO. As for the doped graphene, the Lindhard function is given by,~\cite{Hwang}
\be
	\Pi^{\text{D}}(z\hat q)=\mathcal D(E_F)\left[
	1-\frac{1}{2}\Theta(z-1)(\sqrt{1-\frac{1}{z^2}}+z\arcsin\frac{1}{z}-\frac{\pi z}{2})
	\right]\:,
\ee with which we can deduce that $\Pi^{\text{D}}\approx\mathcal D(E_F)[1-\Theta(z-1)\sqrt{2(z-1)^3}]$ near the singular point $z=1$. $\mathcal D(E_F)$ is the corresponding density of state at Fermi level.

\section{Friedel oscillation}\label{fo}


We now calculate the induced charge density $\delta n(\bf R)$ due to an external charge density $Ze\delta(\bf R)$,~\cite{Guinea}
\be
	 \delta n({\bf R}) = \frac{Ze}{4\pi^2}\int  d{\bf q} \left[
	 \frac{1}{\epsilon({\bf q})}-1\right]
	 e^{i{\bf q}\cdot{\bf R}}\:,
\ee in which the dielectric function given in Ref.~\onlinecite{Hwang} is $\epsilon({\bf q})=1+v_c(q)\Pi(\bf q)$ with $v_c(q)=2\pi e^2/\kappa q$ in the random phase approximation. 
In the limit of large $R$, following Ref.~\onlinecite{Badalyan2}, we employ the method of steepest descent to integrate out the angle between ${\bf q}$ and ${\bf R}$. Assuming the variation of $\Pi(\bf q)$ with respect to orientation of ${\bf q}$ is smooth, the induced charge density is shown to be

\be
	\delta n({\bf R}) \approx Ze\sqrt{\frac{2}{\pi^3R}}\int^{\infty}_0dq\sqrt{q}
	\frac{\cos(qR-\pi/4)}{\epsilon(q\hat R)}\:.
\ee For the case of 2DEG, the singular contribution in $\epsilon(q)^{-1}$ is proportional to $\Theta(q-2k_F)\sqrt{q-2k_F}$, giving rise to, 
\be
	\delta n^{\text{2DEG}}\propto 
	\frac{q_{\text{TF}}k_F^3}{(2k_F+q_{\text{TF}})^2}
	\frac{\cos(2k_FR)}{(2k_FR)^2}
	=\gamma_1(n)\frac{\cos(2k_FR)}{(2k_FR)^2}\:,\label{FO_2DEG}
\ee with the Thomas-Fermi screening vector $q_{\text{TF}}=2\pi e^2D/\kappa$ as defined in Ref.~\onlinecite{DasSarmaRev}. The parameter $\gamma_1$ depending on the carrier density $n$ will be discussed later. The same method is applicable to the case of saddle points in tGB and TCI. For $\hat R\|\hat x$, the leading singular contribution to $\epsilon(q)^{-1}$ is proportional to $\Theta(k_c-q)\sqrt{k_c-q}$, and a similar oscillation is generated,
\be
	\delta n^{\text{vHs}}\propto\frac{q_{\text{TF}}k_c^3}{[\pi k_c+q_{\text{TF}}(2+2\ln{4\Lambda/k_c})]^2}
	\frac{\cos(k_cR)}{(k_cR)^2}
	=\gamma_2(n)\frac{\cos(k_cR)}{(k_cR)^2}\:,\label{FO_sp}
\ee for nonzero $k_c$. Note that an additional term of $1/R^{5/2}$ is produced when the second singular contribution $\propto\Theta(q-k_c)(q-k_c)$ is considered in $\delta n^{\text{vHs}}$. Now an interesting situation arises as $k_c$ is approaching zero, which corresponds to the critical Fermi surface right at the saddle point. Suggested by Eq.~\ref{FO_sp}, one may conclude that the FO disappears when $k_c=0$. From Eq.~\ref{sp_pi}, however, the Lindhard function $\Pi^{\text{vHs}}_0$ does have a singularity at $q_x=0$. Nevertheless, the singularity which as well corresponds to the divergent density of state at saddle point can be removed in the random phase approximation. Namely, the Lindhard function is modified as $\Pi_0^{\text{vHs}}\rightarrow\Pi_0^{\text{RPA}}=\Pi^{\text{vHs}}_0/(1+v_c\Pi^{\text{vHs}}_0)$. It follows that the absence of singularity in $\Pi^{\text{RPA}}$ implies the {\it absence} of FO in the situation we have discussed. The absence of oscillatory term in $\delta n$ was also predicted in intrinsic bilayer graphene in which the response function is a constant.~\cite{Hwang1} Last, when the Fermi surface is close to zero energy and a Dirac point is enclosed, the corresponding FO can be shown to be,
\be
	\delta n^{\text{D}}\propto
	\frac{q^D_{\text{TF}}k_D^3}{(k_D+q^D_{\text{TF}})^2}
	\frac{\cos(k_DR)}{(k_DR)^3}
	=\gamma_3(n)\frac{\cos(k_DR)}{(k_DR)^3}\:,\label{FO_D}
\ee with the energy-dependent screening vector $q^D_{\text{TF}}=2\pi e^2\mathcal D(E_F)/\kappa$. The dependence of $1/R^3$ is the result of the singular contribution of $\Theta(q-2k_F)\sqrt{(q-2k_F)^3}$ in $\Pi^{\text{D}}$.

\section{Discussions}\label{dis}

The previous studies~\cite{Falko,Guinea,Hwang,Herb,Hwang1} have established the fact that the FO's in extrinsic graphene follow that $\delta n\propto R^{-3}$. Suggested by the model band structure in this paper, it is possible to observe such characteristic FO when the Fermi level is near zero energy.  
However, as the Fermi level is raising to higher energy, it is possible to see that $\delta n$ along the principal direction ($\bar\Gamma\bar X_1$ on the surface of TCI, for instance) should evolve to the regime of $R^{-2}$ based on Eq.~\ref{FO_sp}. Therefore, there are two types of FO's at different energies, which is similar to the situation in the surface of topological insulator where hexagonal warping effects are important at higher energy.~\cite{FuHexagonal} Moreover, when the Fermi level is exactly at the saddle point, the oscillation disappears completely, which can serve as a signature of saddle point in the STM measurement. 


The static Lindhard function studied in the paper is also relevant to collective excitations mediated by electrons. Examples include the electron-phonon coupling,~\cite{Phonon2,Phonon} RKKY interaction~\cite{Herb} between magnetic impurities, and screening. A dimensionless ration $q_s\equiv q_{\text{TF}}/k_F$ is an important parameter controlling the strength of quantum screening.~\cite{Hwang1} It has been shown that the Coulomb interaction remains unscreened in graphene due to $q_s$ being a constant, which is in contrast to 2DEG and bilayer graphene where $q_s\propto 1/\sqrt{E_F}$ and the screening is strong in the low density limit.~\cite{Hwang1} Near the van Hove singularity, one can define a similar $q_s=\tilde q_{\text{TF}}/k_c$ with $\tilde q_{\text{TF}}=q_{\text{TF}}\ln(4\lambda^2)$ obtained from the long-wavelength limit of $\Pi^{\text{vHs}}$. We conclude that along the principal axis of hyperbolic Fermi surface, the 2D screening become even stronger $q_s\propto(\ln\frac{E_c}{E-E_{\text{vHs}}})/\sqrt{E-E_{\text{vHs}}}$ when Fermi level is close to the van Hove singularity $E_{\text{vHs}}$ than in the case of 2DEG.  

The detailed structure of oscillating density is included in the parameters $\gamma$'s. For 2DEG, we may write $\gamma_1=\frac{q_{\text{TF}}^2}{q_s(2+q_s)^2}$ in Eq.~\ref{FO_2DEG}, from which $\gamma_1\propto\sqrt{n}$ for $n\gg10^{12}$ $\text{cm}^{-2}$ and $\gamma_1\propto \sqrt{n^3}$ for $n\ll10^{10}$ $\text{cm}^{-2}$ in $n$-GaAs 2DEG.~\cite{DasSarmaRev} In graphene, we may write from Eq.~\ref{FO_D} that $\gamma_3=(q_{\text{TF}}^D)^2/[q_s^D(2+q_s^D)^2]\propto n$ since $q_s^D\approx 3.2$ independent of carrier density $n$.~\cite{DasSarmaRev} Given Eq.~\ref{FO_sp}, we find that $\gamma_2$ has a similar dependence on $n$ with $\gamma_1$ except the presence of the factor $\ln4\lambda^2$ which has a weak dependence on the carrier density.

In conclusion, we study the Friedel oscillation for two-dimensional Dirac materials when the Fermi level is around the van Hove singularity. With approximating the Fermi surface near the saddle point with a hyperbola, we calculate the Lindhard response function $\Pi$ and obtain the induced charge density from the singularity of $\Pi$ using the Lighthill's theorem. The varying Friedel oscillation as the Fermi level is changed can be observed in STM measurements.

\acknowledgments
The author is indebted to Herb Fertig for many insightful discussions. Useful discussions with Arijit Kundu and H.-C. Kao are also acknowledged. This work is supported by Taiwan Ministry of Science and Technology through Grant No.~103-2112-M-003-012-MY3.

\section{Appendix}

The integrals associated with $\Pi$ is proceeded with first changing variable, which yields, 
\be
	\int_0^{\lambda} dy\frac{y^2}{z\pm\sqrt{y^2+1}} = \int_0^{\Theta} d\theta\ \frac{\sec\theta\tan^2\theta}{z\cos\theta\pm1}\:,
\ee with $\Theta=\tan^{-1}\lambda$, and follows with the identity, 
\be
	 \frac{\sec\theta\tan^2\theta}{z\cos\theta\pm1} = \pm\sec\theta\tan^2\theta -z\sec^2\theta \pm z^2\sec\theta \pm \frac{z-z^3}{z\cos\theta+1}\:.
\ee Finally, the formulas below 

\be
	\int dx\sec x = \frac{1}{2}\ln\frac{1+\sin x}{1-\sin x}\:, 
\ee and, 
\bea
	\int \frac{dx}{a+b\cos x} &=& \frac{2}{\sqrt{a^2-b^2}}\tan^{-1}\frac{\sqrt{a^2-b^2}\tan x/2}{a+b}\:,\ [a^2>b^2]\\
	&=& \frac{1}{\sqrt{b^2-a^2}}\ln|\frac{\sqrt{b^2-a^2}\tan x/2+a+b}{\sqrt{b^2-a^2}\tan x/2-a-b}|\:.\ [b^2>a^2]
\eea are employed in obtaining $P(z)$.


\begin{thebibliography}{plain}

\bibitem{vanHove} L. Van Hove, Phys. Rev. {\bf 89}, 1189 (1953).

\bibitem{vHS1} J. Friedel, J. Phys.: Condens. Matter. {\bf 1}, 7757 (1989).

\bibitem{vHS2} R.~S. Markiewicz, J. Phys. Chem. Solids {\bf 58}, 1179 (1997).

\bibitem{vHS3} J. Bouvier and J. Bok, Adv. Condes. Matter Phys. {\bf 2010}, 472636 (2010).

\bibitem{Tsuei} C.~C. Tsuei, D.~M. Newns, C.~C. Chi, and P.~C. Pattnaik, Phys. Rev. Lett. {\bf 65}, 2724 (1990).

\bibitem{highTC} P.~.C. Pattnaik, C.~L. Kane, D.~M. Newns, and C.~C. Tsuei, Phys. Rev. B {\bf 45}, 5714 (1992).

\bibitem{dopedG} J.~L. McChesney, A. Bostwick, T. Ohta, T. Seyller, K. Horn, J. Gonzalez, E. Rotenberg, Phys. Rev. Lett. {\bf 104}, 136803 (2010).

\bibitem{Gonzalez0} J. Gonzalez, Phys. Rev. B {\bf 78}. 205431 (2008).

\bibitem{Nandkishore}  R. Nandkishore, L.~S. Levitov, and A.~V. Chubukov, Nature Phys. {\bf 8}, 158 (2012).

\bibitem{tBG0} G. Li, A. Luican, J.~M.~B. Lopes dos Santos, A.~H. Castro Neto, A. Reina, J. Kong, and E.~Y. Andrei, Nat. Phys. {\bf 6}, 109 (2010).

\bibitem{TCI1} T.~H. Hsieh, H. Lin, J. Liu, W. Duan, A. Bansil, and L. Fu, Nature Commun. {\bf 3}, 982 (2012). 


\bibitem{P0} H. Liu, A.~T. Neal, Z. Zhu, Z. Luo, X. Xu, D. Tomanek, and P.~D. Ye, ACS Nano {\bf 8}, 4033 (2014).

\bibitem{P1} A.~S. Rodin, A. Carvalho, and A.~H. Castro Neto, Phys. Rev. Lett. {\bf 112}, 176801 (2014).

\bibitem{HLin} A. Ziletti, S.-M. Huang, D.~F. Coker, and H. Lin, Phys. Rev. B {\bf 92}, 085423 (2015).

\bibitem{tBGa} J.~M.~B. Lopes dos Santos, N.~M.~R. Peres, and A.~H. Castro Neto, Phys. Rev. Lett. {\bf 99}, 256802 (2007).

\bibitem{Shallcross} S. Shallcross, S. Sharma, and O. A. Pankratov, Phys. Rev. Lett. {\bf 101}, 056803 (2008).

\bibitem{Brey} T. Stauber, P. San-Jose, and L. Brey, New J. Phys. {\bf 15}, 113050 (2013).

\bibitem{tBG1} R. de Gail, M.~O. Goerbig, F. Guinea, G. Montambaux, and A.~H. Castro Neto, Phys. Rev. B {\bf 84}, 045436 (2011).

\bibitem{tBGKorea} M.-Y. Choi, Y.-H. Hyun, and Y. Kim, Phys. Rev. B {\bf 84}, 195437 (2011).

\bibitem{Bis} R. Bistritzer and A.~H. MacDonald, Phys. Rev. B {\bf 84}, 035440 (2011).

\bibitem{tBGMoon} P. Moon and M. Koshino, Phys. Rev. B {\bf 85}, 195458 (2012).

\bibitem{tBGLu1} C.-K. Lu and H.~A. Fertig, Phys. Rev. B {\bf 89}, 085408 (2014).

\bibitem{tBGLu2} C.-K. Lu and H.~A. Fertig, Phys. Rev. B {\bf 90}, 115436 (2014).

\bibitem{FuTCI_PRB} M. Serbyn and L. Fu, Phys. Rev. B {\bf 90}, 035402 (2014).

\bibitem{STM} A. Luican-Mayer, M. Kharitonov, G. Li, C.-P. Lu, I. Skachko, A.-M.~B. Goncalves, K. Watanabe, T. Taniguchi, and E.~Y. Andrei, Phys. Rev. Lett. {\bf 112}, 036804 (2014).

\bibitem{Friedel} J. Friedel, Phil. Mag. {\bf 43}, 153 (1952).

\bibitem{Stern} F. Stern, Phys. Rev. Lett. {\bf 18}, 546 (1967).

\bibitem{Ando} T. Ando, J. Phys. Soc. Jpn. {\bf 75}, 074716 (2006).

\bibitem{DHLin} D.-H. Lin, Phys. Rev. A {\bf 72}, 012701 (2005).

\bibitem{Falko} V.~V. Cheianov and V.~I. Fal'ko, Phys. Rev. Lett. {\bf 97}, 226801 (2006).

\bibitem{Guinea} B. Wunsch, T. Stauber, F. Sols, and F. Guinea, New J. Phys. {\bf 8}, 318 (2006).

\bibitem{Hwang} E.~H. Hwang and S. Das Sarma, Phys. Rev. B {\bf 75}, 205418 (2007).

\bibitem{Herb} L. Brey, H.~A. Fertig, S. Das Sarma, Phys. Rev. Lett. {\bf 99}, 116802 (2007).

\bibitem{Kohn} W. Kohn, Phys. Rev. Lett. {\bf 2}, 393 (1959).

\bibitem{lighthill} M.~J. Lighthill, {\it Fourier Analysis and Generalised Functions} (Cambridge University Press, New York, 1958).


\bibitem{nematic2} R. Nandkishore and L. Levitov, Phys. Rev. Lett. {\bf 107}, 097402 (2011).

\bibitem{Hwang1} E.~H. Hwang and S. Das Sarma, Phys. Rev. Lett. {\bf 101}, 156802 (2008).



\bibitem{Gonzalez1} J. Gonzalez, F. Guinea, and M.~A.~H. Vozmediano, Europhys. Lett. {\bf 34}, 711 (1996).



 








\bibitem{Badalyan2} S.~M. Badalyan, A. Matos-Abiague, G. Vignale, and J. Fabin, Phys. Rev. B {\bf 81}, 205314 (2010).




\bibitem{DasSarmaRev} S. Das Sarma, S. Adam, E.~H. Hwang, and E. Rossi, Rev. Mod. Phys. {\bf 83}, 407 (2011).

\bibitem{FuHexagonal} L. Fu, Phys. Rev. Lett. {\bf 103}, 266801 (2009).


\bibitem{Phonon2} C.-H. Park, F. Giustino, J.~L. McChesney, A. Bostwick, T. Ohta, E. Rotenberg, M.~L. Cohen, and S.~G. Louie, Phys. Rev. B {\bf 77}, 113410 (2008).


\bibitem{Phonon} A. Politano, F. de Juan, G. Chiarello, and H.~A. Fertig, Phys. Rev. Lett. {\bf 115}, 075504 (2015).
















\end{thebibliography}
\end{document}